\documentclass[fontsize=11pt,a4paper]{article}
\usepackage[T1]{fontenc}
\usepackage[utf8]{inputenc}
\usepackage[english]{babel}
\usepackage{graphicx}
\usepackage[a4paper]{geometry}
\usepackage{lmodern}
\usepackage{lineno}
\usepackage{color}

\ifdefined\REVIEW
   \linenumbers
   
\fi

\usepackage[hidelinks]{hyperref}
\usepackage{amsmath}

\usepackage{soul}
\title{High-Performance Computing in Battery Development: From Pore Scale to Continuum}

\author{
        Benjamin Kellers\textsuperscript{1,2,}%
        \footnote{Corresponding author, Email: \href{mailto://benjamin.kellers@dlr.de}{benjamin.kellers@dlr.de}}
            \and
        Martin P.\ Lautenschlaeger\textsuperscript{1,2}
            \and
        Julius Weinmiller\textsuperscript{1,2}
        \and
        Lukas Krumbein\textsuperscript{1,2}
            \and
        Simon Hein\textsuperscript{1,2}
        \and
        Timo Danner\textsuperscript{1,2}
            \and
        Arnulf Latz\textsuperscript{1,2,3}
        \vspace{0.5em}
        \and
        \small
        \textsuperscript{1}~German Aerospace Center (DLR), Institute of Engineering Thermodymanics, 89081 Ulm, Germany.\\
        \small
        \textsuperscript{2}~Helmholtz Institute Ulm for Electrochemical Energy Storage (HIU), 89081 Ulm, Germany.\\
        \small
        \textsuperscript{3}~Ulm University, Institute of Electrochemistry, 89081 Ulm, Germany.%
}

\date{%
}
\begin{document}
\maketitle

%
%

\begin{abstract}
An application for high-performance computing (HPC) is shown that is relevant in the field of battery development. Simulations of electrolyte wetting and flow are conducted using pore network models (PNM) and the lattice Boltzmann method (LBM), while electrochemical simulations are conducted using the tool BEST. All aforementioned software packages show an appropriate scaling behavior. A workflow for optimizing battery performance by improving the filling of battery components is presented. A special focus is given to the unwanted side effect of gas entrapment encountered during filling. It is also known to adversely affect the electrochemical performance of batteries and can be partially prevented by appropriate microstructure design such as electrode perforation. 

\end{abstract}

\section{Introduction}\label{introduction}

Lithium-ion batteries play a major role in electronics with applications ranging from mobile phones to electric vehicles. Although the technology could be considered mature, there is still room for improvement in several aspects of, e.g., the production. One of them is the filling with liquid electrolyte. This process is time-critical, cost-intensive and known to have a significant influence on the electrochemical performance of batteries. 

The underlying reason is that battery components typically consist of porous media to maximize reactive surface areas and the energy density in a given volume. However, the porosity including pore morphology and topology as well as differences in wettability of electrolyte and ambient air induce issues during the filling. They can result in partial gas entrapment, i.e. incomplete electrolyte saturation. Thus, only parts of the surface of the active material in the electrode are submersed with electrolyte, while the rest is considered electrochemically inactive. Moreover, the entrapped air is an insulator for ionic and electric conductivity. Technically speaking, the effective tortuosity, i.e. the length of the shortest path between two opposite sides of the structure, is increased. This finally leads to a decrease of the battery performance. 

Capturing these phenomena through experiments is challenging and sometimes even impossible. Thus, theoretical modeling and simulations are instead applied here. Electrolyte wetting is simulated using pore network models (PNM) and the lattice Boltzmann method (LBM). The software tool BEST is used to predict the electrochemical performance. All tools are connected in a workflow. A critical aspect of this study is the multi-criteria optimization. Structural parameters and material properties which improve the filling might still worsen the battery performance. Thus, creating a feedback loop in the process optimization is key for the proposed workflow.

The present study is structured as follows: In Sec.\,\ref{setup} an overview of the workflow is given and the computational methods and the software packages are introduced. Simulation results are presented and discussed in Sec.\,\ref{results}. Information on HPC scalability of the software tools are given in Sec.\,\ref{hpc}. Finally, conclusions are drawn in Sec.\,\ref{conclusions}, where also a short outlook is given.

\section{Workflow and software packages}\label{setup}

\begin{figure}[htb] 
  \begin{center}
    \includegraphics[scale=0.5]{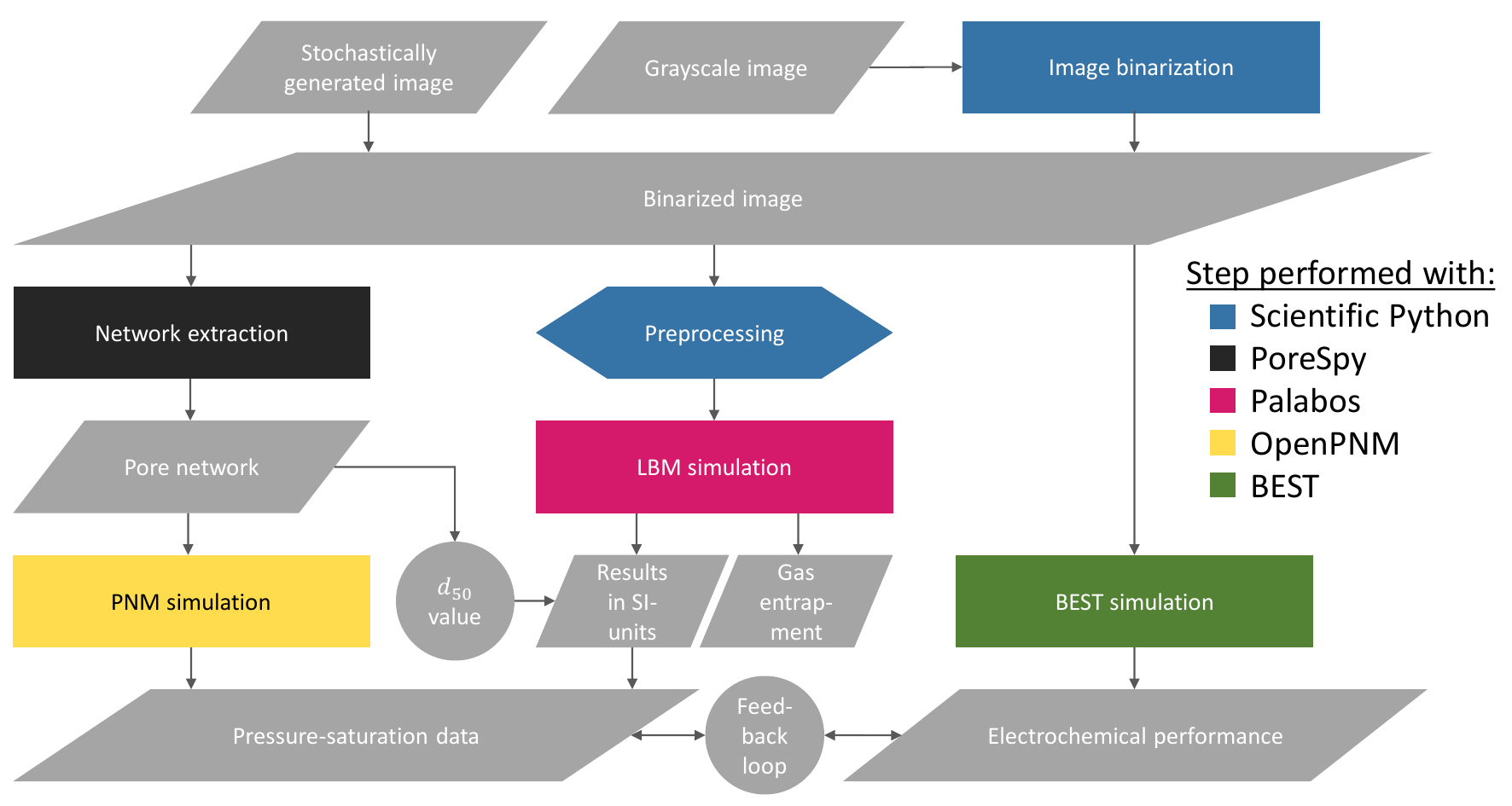}
    \caption{Flowchart of the workflow. The data flow and the relationships between the different steps are illustrated. The boxes are colored with respect to the software used. Gray labels depict steps that are not directly related to any specific software tool.}
    \label{fig:workflow}
  \end{center}
\end{figure}

The workflow is schematically shown in Fig.\,\ref{fig:workflow}. It is described in the following, going from top to bottom and left to right.

Since pore-scale simulations are applied, 3D images of porous battery electrodes or separators are required. In this work, lithium-ion battery cathodes are used as an example. The corresponding structural data is either virtually generated using physically-based algorithms\,\cite{Westhoff2018} or comes as grayscale images from tomography techniques (FIB-SEM or micro-CT) of real components which requires an additional binarization step in the workflow. The latter, for brevity, is however not described here. 

In the workflow, as mentioned before, three different software tools and also simulation techniques are used and connected. They are depicted by the three vertical strands in Fig.\,\ref{fig:workflow}. From the left to the right, these correspond to PNM using OpenPNM\,\cite{Gostick2016}, LBM using Palabos\,\cite{Latt2021}, and electrochemical simulations using BEST\,\cite{best}, i.e.\ a finite volume implementation of Li-ion battery models developed by our group\,\cite{latz}.

\begin{figure}[htb] 
  \begin{center}
    \includegraphics[scale=0.5]{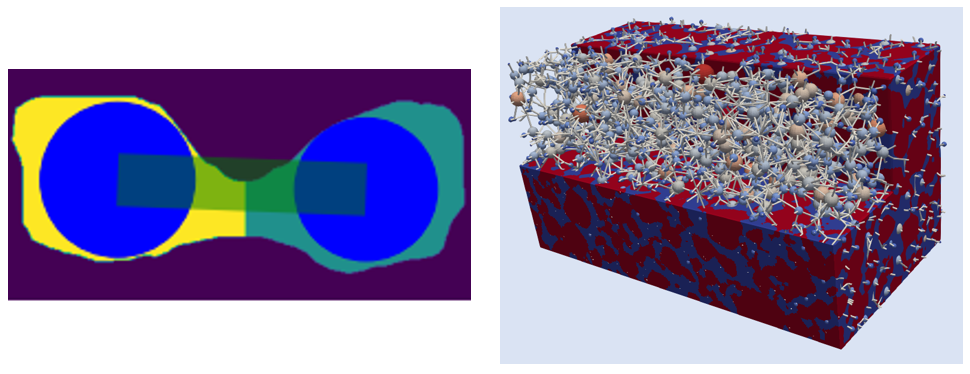}
    \caption{Network extraction from structurally-resolved image data. Left: 2D pore space segmented into two pores (yellow and green). The disk sizes (blue) correspond to the inscribed diameters of the pores and are connected by a throat (opaque rectangle). Right: Extracted full network embedded into the corresponding 3D image of a battery cathode. The pore color corresponds to the pore volume. The throat diameters are depicted for topological reasons only and do not relate to the physical size.}
    \label{fig:network}
  \end{center}
\end{figure}

For the PNMs, based on the binarized image data the pore network information is extracted. Therefore, the open-source Python library PoreSpy\,\cite{Gostick2019} is used. It reconstructs the geometrical and topological properties of the structurally-resolved pore space using simplified geometries, which typically means spherical pores connected by cylindrical throats (cf. Fig.\,\ref{fig:network}). The underlying procedure is based on a corrected watershed algorithm to segment the image data. In the current work, the so-called SNOW-algorithm\,\cite{Gostick2017} is used to also avoid oversegmentation; an issue often observed when segmenting energy materials. Fig.\,\ref{fig:network} shows examples of such extracted networks. On the left-hand side, a simplified 2D case consisting of only two pores is schematically depicted. On the right-hand side, a full network for a 3D battery electrode structure consisting of several thousand pores is given. After the network extraction, the network properties, such as for example the median pore diameter $d_{\text{50}}$ can be computed. It can also be used to determine other physical properties of interest. The method is computationally very efficient, since only analytical expressions are solved for pores and throats. In the current work, a percolation into a pore network is simulated to determine the pressure-saturation behavior during the filling process. 

For the LBM simulations an additional preprocessing step of the binarized image data is needed in which solid interfaces are differentiated from the solid bulk. This helps to speed up the simulations. The preprocessing is done using the Python libraries NumPy and SciPy. The filling process itself is modeled as a two-phase flow using the Shan-Chen method\,\cite{Shan1993}. Palabos\,\cite{Latt2021} is used for the simulations from which also pressure-saturation curves are determined. They are used for validating the results determined using PNM. Moreover, the local distribution of entrapped residual gas is finally analyzed. For further details, the reader is referred to Ref.\,\cite{Lauten2022a}.

With respect to Fig.\,\ref{fig:workflow}, in the workflow, LBM and PNM are connected through an information exchange step. It is necessary to convert the LBM-specific unit system to physical units. Therefore, geometrical information about the pore space is needed, i.e.\ the median pore diameter $d_{\text{50}}$ is used in the present study. In exchange, one pressure-saturation data set determined using LBM is used for calibration of the PNM. Details regarding this calibration and necessary corrections in the PNM will be discussed in a future publication of our group. The calibrated PNM can then be used to cover huge parameter spaces in percolation simulations to gain understanding of the influence of different input parameters on the pressure-saturation behavior. Since the PNM computations are extremely lightweight, this speeds up the optimization of the filling process by several orders of magnitude. 

The binarized image data of the porous electrodes also serves as input for the electrochemical simulations. The corresponding domain consists of the electrode, a separator, and a counter-electrode. On this geometrical basis a set of coupled partial differential and algebraic equations consisting of mass, charge and energy balances is solved. The parametrization of the conservation equations is done through dedicated experiments or using literature data.

Finally, simulation results regarding pressure-saturation data and electrochemical performance are considered in a feedback loop, that aims to improve both aspects while minimizing negative effects on each of them.

\section{Results and discussion}\label{results}

In the following, a selection of representative results are discussed only. For the sake of completeness the reader is referred to related publications from our group\,\cite{Lauten2022a,Lauten2022b,DeLauri2021}.

\begin{figure}[htb] 
  \begin{center}
    \includegraphics[scale=0.5]{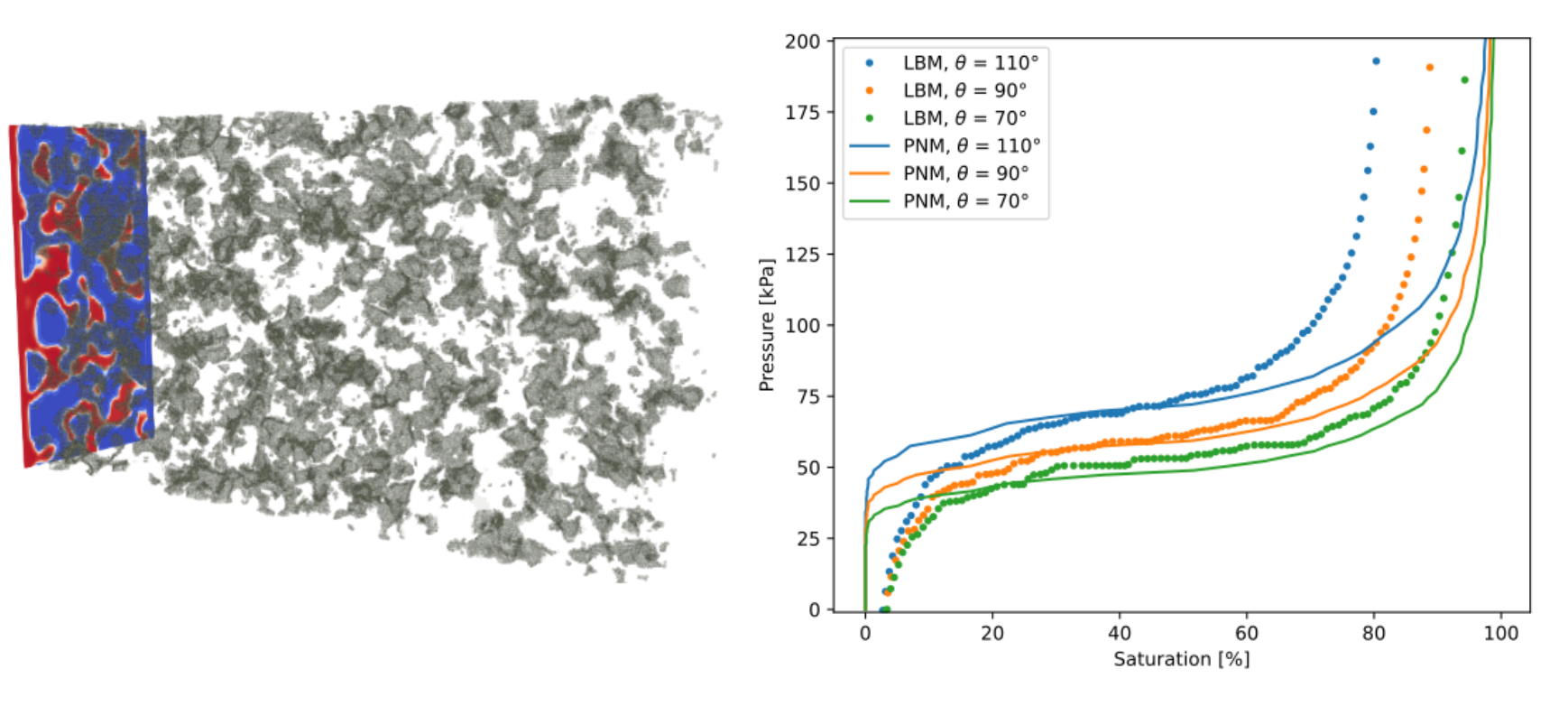}
    \caption{Simulation results for electrolyte filling using LBM and PNM. Left: Entrapped gas (gray) in the pore space at the end of the filling process. Electrolyte and active material are fully transparent. Right: Pressure-saturation curves determined using LBM (dots) and PNM (lines) simulations for a selection of contact angles $\theta$.}
    \label{fig:LBM}
  \end{center}
\end{figure}

Fig.\,\ref{fig:LBM} illustrates the final state of an LBM filling simulation and the pressure-saturation curves corresponding to the simulation setup. It can be seen from Fig.\,\ref{fig:LBM} that a relatively large amount of entrapped gas remains inside the porous structure. This is also obvious from the LBM pressure-saturation curves, which show that the final degree of saturation for different contact angles between electrolyte and active material ranges between 75\,\% and 95\,\%. The PNM does not yet model gas entrapment, such that all PNM simulations reach 100\,\% saturation in this study.

\begin{figure}[htb] 
  \begin{center}
    \includegraphics[scale=1.0]{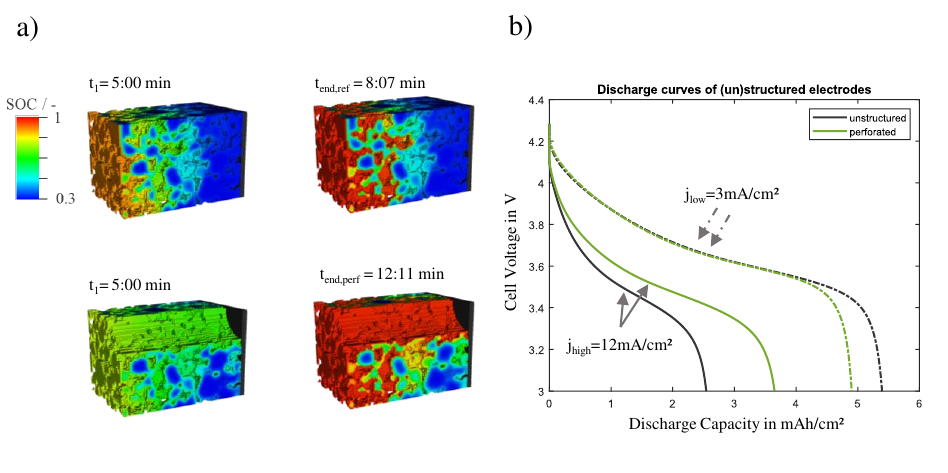}
    \caption{Electrochemical aspects. a) SOC distribution of lithium in unstructured electrode (top) and structured electrode (bottom). The structuring improves the homogeneity of the SOC distribution. b) Electrochemical performance for operation currents of $j\geq 6\text{mA/cm}^2$.}
    \label{fig:perf_electrode}
  \end{center}
\end{figure}

Fig.\,\ref{fig:perf_electrode} shows results of the electrochemical simulations. Two types of electrodes are studied: (1) unstructured electrodes and (2) structured electrodes with improved wettability. In experiments the structure is realized, by selectively removing electrode material in a cylindrical hole pattern with a short-pulsed laser. This procedure is mainly applied to thick electrodes and known to improve rate capability, cycle stability, tendency to lithium plating\,\cite{DeLauri2021} as well as the wettability and electrolyte saturation\,\cite{Kleefoot2022}. 

Fig.\,\ref{fig:perf_electrode}a) shows the state of charge (SOC) distribution within unit cells of the unstructured and structured electrodes at different times. The structuring by perforation improves the wettability and thereby provides additional transport channels for lithium ions with low tortuosity. Both effects are able to improve the utilization of the active material illustrated by the increase in red areas shown in Fig.\,\ref{fig:perf_electrode}a).  Fig.\,\ref{fig:perf_electrode}b) shows the accessible capacity of an electrode with $6\,\text{mAh/cm}^2$ theoretical capacity: The perforation improves the electrochemical performance for high operating currents $j$.

\section{Necessity and benefits of using HPC infrastructures}\label{hpc}

Multiple Python libraries and some custom in-house code are involved in the described workflow. More specifically, the scientific stack of Python consisting of NumPy and SciPy as well as Scikit-Image\,\cite{skimage} were used. These partly rely on precompiled C-code which often comes with multi-threading capabilities. However, they do not utilize MPI optimization out-of-the-box. Thus, all steps in the aforementioned workflow, for which Python was used, were run on a single CPU only, though making use of multiple cores or threads. The same applies for the network extraction using PoreSpy as well as the PNM simulations using OpenPNM, as both are based on the scientific stack of Python, too.

\begin{figure}[htb] 
  \begin{center}
    \includegraphics[scale=0.62]{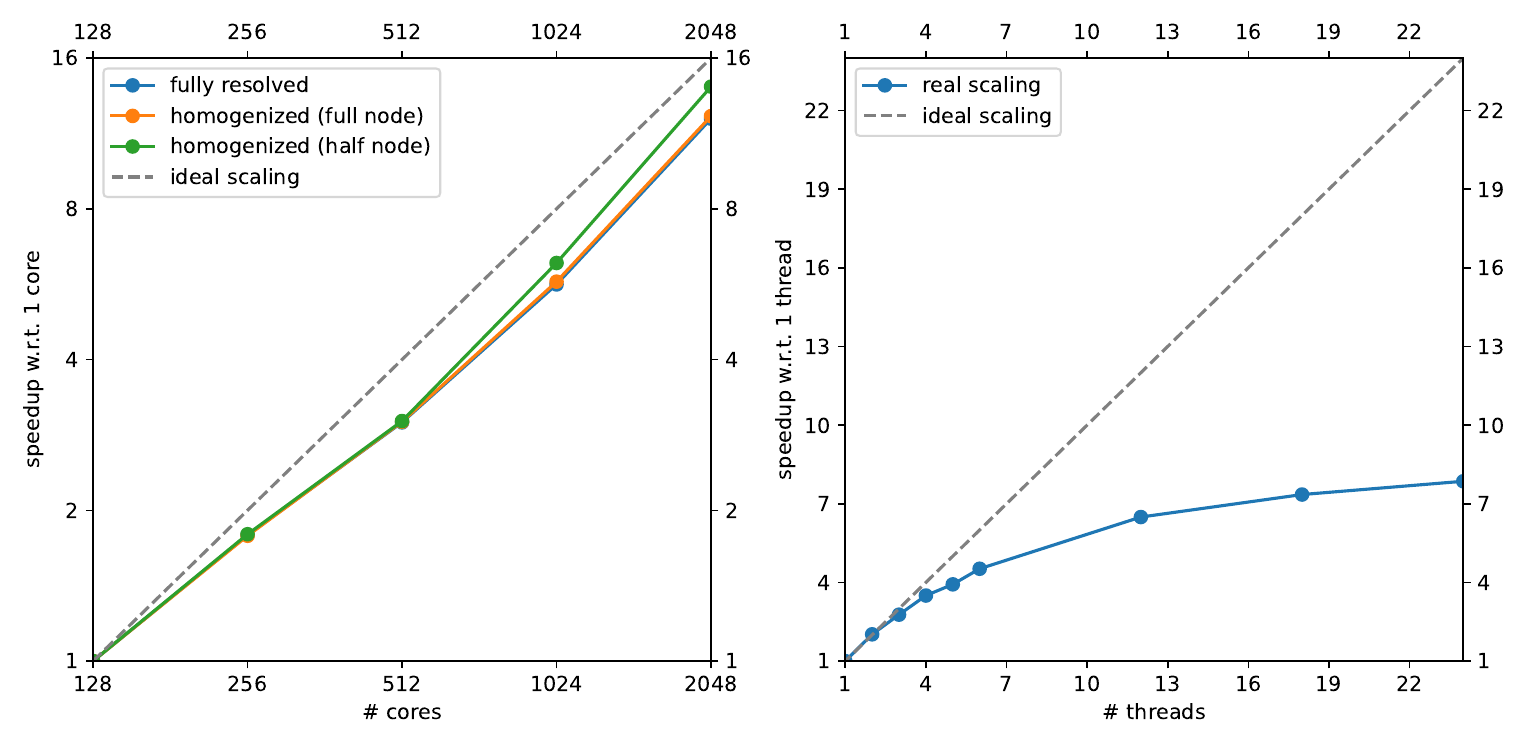}
    \caption{Performance Test. Left: Scaling of fully resolved and homogenized LBM electrolyte filling simulations using Palabos. Right: Scaling of a matrix multiplication in Python using NumPy in a multi-threaded environment to illustrate the scaling of non-MPI computations in the workflow.}
    \label{fig:scaling}
  \end{center}
\end{figure}

Therefore, performance characteristics of precompiled Python C-code are shown in Fig.\,\ref{fig:scaling} using the example of a matrix multiplication with NumPy. However, also with respect to PoreSpy, full use of the HPC environment was made regarding the usage of RAM size and memory bandwidth. Its network extraction algorithm can be very RAM-demanding, since multiple copies of the original image are required for the necessary computations. Palabos, on the other hand, hugely benefits from an HPC environment, since it is fully MPI optimized. The models used require only local information which is extremely favorable for multi-core parallelization. Fig.\,\ref{fig:scaling} shows the huge difference the MPI optimization makes when comparing to regular multi-threaded scaling.

\section{Conclusions and Outlook}\label{conclusions}

This study shows the potential power of combining different simulation tools into a workflow by information exchange and feedback loops. Often in process optimization, parameters have a negative effect on each other such that trade-offs are necessary to improve the process as a whole. This is showcased here by the interdependence of electrolyte filling and electrochemical performance of batteries. Another aspect highlighted in this study, is the fact that software which is not HPC-optimized per se, can still benefit from HPC environments. Additionally, coupling of low and high fidelity models result in the necessary efficiency to use such a workflow on large parameter spaces.

As part of our future work, this study will be extended in two directions. First, by performing electrochemical simulations on partially saturated electrodes determined using LBM simulations. Second, by simulating the effect of structured or laser-perforated electrodes on the electrolyte filling. 

\section*{Acknowledgements} 

We gratefully acknowledge financial support from the European Union's Horizon 2020 Research and Innovation Programme within the project "\textbf{DEFACTO}" [grant number 875247] for parts of the LBM and PNM activities of our research as well as from the ‘Bundesministerium für Bildung und Forschung’ within the project \textbf{HiStructures} under the reference No.\ 03XP0243E. This work also contributes to the research performed at \textbf{CELEST} (Center for Electrochemical Energy Storage Ulm-Karlsruhe). The simulations were carried out on the Hawk at the High Performance Computing Center Stuttgart (\textbf{HLRS}) [grant \textbf{LaBoRESys}], and on \textbf{JUSTUS 2} at Ulm University [grant INST 40/467-1 FUGG]. 


\end{document}